\documentclass[utf8]{FrontiersinHarvard} % for articles in journals using the Harvard Referencing Style (Author-Date), for Frontiers Reference Styles by Journal: https://zendesk.frontiersin.org/hc/en-us/articles/360017860337-Frontiers-Reference-Styles-by-Journal
\usepackage{url,hyperref,lineno,microtype,subcaption}
\usepackage[onehalfspacing]{setspace}
\usepackage{amsmath}
\def\keyFont{\fontsize{8}{11}\helveticabold }
\def\firstAuthorLast{J. Hartke} %use et al only if is more than 1 author
\def\Authors{Johanna Hartke\,$^{1,2,3*}$}
\def\Address{$^{1}$ Finnish Centre for Astronomy with ESO (FINCA), University of Turku, Turku, Finland,\\
$^{2}$ Tuorla Observatory, Department of Physics and Astronomy, University of Turku, Turku, Finland,\\
$^{3}$ Turku Collegium for Science, Medicine and Technology (TCSMT), University of Turku, Turku, Finland}
\def\corrAuthor{Johanna Hartke}

\def\corrEmail{johanna.hartke@utu.fi}

\usepackage{natbib}
\DeclareRobustCommand{\ion}[2]{\textup{#1\,\textsc{\lowercase{#2}}}}

\begin{document}
\onecolumn
\firstpage{1}

\title[PNe as stellar population tracers]{Planetary nebulae as tracers of accreted stellar populations in massive galaxies in groups and clusters} 

\author[\firstAuthorLast ]{\Authors} %This field will be automatically populated
\address{} %This field will be automatically populated
\correspondance{} %This field will be automatically populated

\extraAuth{}% If there are more than 1 corresponding author, comment this line and uncomment the next one.
%\extraAuth{corresponding Author2 \\ Laboratory X2, Institute X2, Department X2, Organization X2, Street X2, City X2 , State XX2 (only USA, Canada and Australia), Zip Code2, X2 Country X2, email2@uni2.edu}

\maketitle

\begin{abstract}

%%% Leave the Abstract empty if your article does not require one, please see the Summary Table for full details.
%%% Follow IMRAD: introduction, methods, results, and discussion
% Introduction
Planetary nebulae (PNe) are valuable spatial and kinematic tracers of nearby galaxies. In this mini-review, I focus on their role in tracing the halo and intra-cluster/intra-group light assembly in groups and clusters of galaxies within 100~Mpc and, in particular, the link between characteristic PN metrics such as the $\alpha$-parameter and the PN luminosity function and changes from the underlying in-situ to ex-situ stellar populations. These results from nearby groups and clusters are placed into context with integral-field spectroscopic surveys of galaxies, which allow the co-spatial measurement of PN and stellar population properties. I provide an outlook on upcoming instrumentation that will provide new opportunities for the study of extragalactic PN populations. I address the challenges of reconciling observations of extragalactic PN populations with predictions from stellar evolution models and how revised late-stellar-evolution models have alleviated some of the tensions between observations and theory.

\tiny
 \keyFont{ \section{Keywords:} planetary nebulae: general, galaxies: clusters, galaxies: groups, galaxies; elliptical and lenticular, cD, galaxies: halos} %All article types: you may provide up to 8 keywords; at least 5 are mandatory.
\end{abstract}

%TC:endignore
\section{Introduction}
%- what are PNe? 
Planetary nebulae (PNe) occur the very late evolutionary stages of low-mass (initial masses between $1$ and $8\,M_\odot$) stars. The PN phase is considered part of the post-asymptotic giant branch (post-AGB) phase. The spectra of PNe are characterised by the absence of a stellar continuum and strong nebular emission lines. It is predicted that up to 15\% of the radiation of their central stars is re-emitted in a single line: the [\ion{O}{iii}] $\lambda 5007\textup{~\AA}$ line \citep{dopita_1992ApJ...389...27D}. This line can be used for the identification and radial velocity measurement of \textit{individual} PNe in galaxies at distances as far as 100\,Mpc \citep{gerhard_2005ApJ...621L..93G}, making them excellent tracers of the spatial distribution and kinematics of galaxies and their haloes. 

Extragalactic PN populations are typically characterised with two metrics: the PN luminosity function (PNLF) and the luminosity-specific PN number, $\alpha$-parameter for short. The PNLF describes the $m_{5007}$ magnitude distribution, where $m_{5007}$ is the magnitude corresponding to the flux of the [\ion{O}{iii}] $\lambda 5007\textup{~\AA}$ line \citep{jacoby_1989ApJ...339...39J}. Empirically, the PNLF is near-universal in early- and late-type galaxies, described by the following relation \citep{ciardullo_1989ApJ...339...53C}:
\begin{equation}
    N_\mathrm{PN}(m_{5007}) \propto \left(1 - \mathrm{e}^{3(M^\star - m_{5007} + \mu)}\right)\mathrm{e}^{c_2 (m_{5007} - \mu)},
    \label{eq:PNLF}
\end{equation}
where $M^{\star}$ denotes the absolute magnitude of the bright cut-off, $\mu$ the distance modulus, and $c_2$ the faint-end slope.
To first order, the absolute bright cut-off magnitude has been found to be invariant at $M^{\star} = -4.5$ \citep{ciardullo_2022}, facilitating the use of the PNLF as a secondary distance indicator by determining the apparent bright cut-off magnitude $m^{\star}$ and thus the distance modulus $\mu$. While well-established observationally, the invariant bright cut-off cannot be easily reconciled with predictions from stellar evolution models, in particular for early-type galaxies with old stellar populations that were predicted to have much fainter bright cut-off magnitudes than observed \citep[e.g.][]{marigo_2004A&A...423..995M, ciardullo_2022} until the advent of revised post-AGB stellar evolution models \citep{millerbertolami_2016A&A...588A..25M, gesicki_2018}. 
The second term of eq.~\eqref{eq:PNLF} was originally introduced with the motivation of describing the fading brightness of a gas cloud around the PN central star \citep{henize_1963ApJ...137..747H}, with the slope fixed to $c_2 = 0.307$ based on observations of M31 and other Local Group galaxies \citep{ciardullo_1989ApJ...339...53C}. 

By integrating the PNLF up to a limiting magnitude $m_\mathrm{lim}$, one obtains the total number of PNe, $N_{\rm PN}$, which can be related to the bolometric luminosity $L_\mathrm{bol}$ of the underlying stellar population via the $\alpha$-parameter\footnote{Not to be confused with the $\alpha$-abundance of elements.} \citep{jacoby_1980ApJS...42....1J}:
\begin{equation}
    \label{eq:alpha}
    \alpha = \frac{N_\mathrm{PN}}{L_\mathrm{bol}} = \mathcal{B}\tau_{\rm PN},
\end{equation}
where $\mathcal{B}$ is the specific evolutionary flux\footnote{The rate of stars that evolve to the post-main sequence stages, normalised per unit light of the stellar population \citep[][]{renziniGlobalPropertiesStellar1986}.} and $\tau_{\rm PN}$ the visibility lifetime of PNe.
Accounting for incompleteness effects, and assuming that the bolometric luminosity $L_\mathrm{bol}$ of the host galaxy can be derived from its broadband photometry with a bolometric correction \citep{buzzoni_2006MNRAS.368..877B}, the $\alpha$-parameter can be straightforwardly obtained from PN surveys. 
For a chemically homogeneous, coeval stellar population in the single stellar population (SSP) framework, the $\alpha$-parameter can also be related to the visibility lifetime $\tau_{\rm PN}$ \citep{buzzoni_1989ApJS...71..817B, buzzoni_2006MNRAS.368..877B}. 
Assuming that the specific evolutionary flux $\mathcal{B}$ does not significantly vary between SSPs with different properties such as metallicity or initial mass function \citep{renziniGlobalPropertiesStellar1986}, the $\alpha$-parameter can thus be interpreted as a direct proxy for the PN visibility lifetime.

Typical values of $\alpha$ are calculated within a limiting magnitude $m_\mathrm{lim} = m^{\star} + 2.5$ and are of the order of $\sim10^{-8}$ to $10^{-9}\,L_{\rm \odot,bol}^{-1}$. If the PN survey is shallower, the $\alpha$-parameter can be extrapolated:
\begin{equation}
    \alpha_{2.5} = \rho_{m_\mathrm{lim}} \alpha_{m_\mathrm{lim} - m^{\star}}.
\end{equation}
The extrapolation factor $\rho_{m_\mathrm{lim}}$ can be obtained via integration of the PNLF (eq.~\eqref{eq:PNLF}):
\begin{equation}
    \rho_{m_\mathrm{lim}} = \frac{\int_{m^{\star}}^{m^{\star}+2.5} N(m_{5007})\mathrm{d}m_{5007}}{\int_{m^{\star}}^{m_\mathrm{lim}} N(m_{5007})\mathrm{d}m_{5007}}.
\end{equation}
In this case, the $\alpha$-parameter can only be used to provide an estimate of the PN visibility lifetime within the limiting magnitude of the survey. 

In the past, the primary target of PN surveys for galaxy kinematics have been early-type galaxies \citep[e.g.][]{arnaboldi_1994Msngr..76...40A, hui_1995ApJ...449..592H, coccato_2009MNRAS.394.1249C, cortesi_2013A&A...549A.115C, pulsoni_2018A&A...618A..94P}, for the simple reason that their identification is facilitated by the lack of other strong [\ion{O}{iii}] emitters (such as supernova remnants and \ion{H}{ii} regions) in old stellar populations compared to late-type galaxies with younger stellar populations. Historically, surveys for PNe in galaxies made use of the on-off band technique, where the on band image is taken with a narrow-band [\ion{O}{iii}] filter, and the off band through an adjacent broad-band filter, for example the $V$- or $g$-bands \citep[e.g.][]{ford_1973ApJ...183L..73F, ford_1975ApJ...202..365F}. The detected PNe candidates are then followed up with spectrographs with high multiplexing capabilities to measure their line-of-sight velocity and identify remaining PN mimics such as Lyman-$\alpha$ emitters at high redshift. Counter-dispersed imaging, a variant of slitless spectroscopy, was developed to remove the need for a two-step process, allowing the detection and velocity measurement in one observation with the custom-built Planetary Nebula Spectrograph \citep[PN.S,][]{douglas_1999MNRAS.307..190D, douglas_2002PASP..114.1234D}. However, due to the lack of full spectral information, this technique was predominantly used in early-type galaxies to limit the contamination from other [\ion{O}{iii}] emitters until a H$\alpha$-arm was installed at the PN.S \citep{2018MNRAS.476.1909A, 2021MNRAS.500.3579A}. 

These classical methods excel at detecting PNe in large areas, spanning several hundreds of square-arcseconds, corresponding to multiple effective radii of galaxies in the local Universe, allowing to map the kinematic transition from in-situ to ex-situ stellar haloes \citep[e.g.][]{pulsoni_2018A&A...618A..94P} as well as from haloes to the surrounding intra-cluster and intra-group light (ICL and IGL, respectively) as recently reviewed by \citet[][and references therein]{arnaboldiKinematicsDiffuseIntragroup2022}. In this review, I want to address how these kinematic transitions coincide with observed changes in the PN population properties, how these PN population properties can be linked to the characteristics of the underlying stellar populations, and how both new observing techniques and numerical stellar evolution models are necessary to contextualise these results.

This review is organised as follows: in Section~\ref{sec:obs}, I review recent observational results concerning the link between PN- and stellar populations based on PN-surveys carried out with classical methods and new insights from integral-field spectroscopic surveys, and provide an outlook on new facilities and opportunities. In Section~\ref{sec:models}, I address new developments in modelling PN populations in the context of updated post-AGB stellar evolution models and numerical models of galaxy evolution. I conclude this review in Section~\ref{sec:conclusion}.
 
\section{Characterisation of PN populations from observations}
\label{sec:obs}
\subsection{Classical PN surveys}
\label{ssec:classical}
% PNLF
Surveys for PNe with classical techniques, such as `on-off' imaging and slitless spectroscopy, provided first constraints on the link between stellar-population properties and the $\alpha$- and PNLF-parameters. \citet{ciardullo_2002} suggested an oxygen-abundance-dependent PNLF bright cut-off $M^{\star}$ based on a small sample of a dozen galaxies with known Cepheid distances. \citet{buzzoni_2006MNRAS.368..877B} investigated the variation of the $\alpha$-parameter with the underlying stellar population in the context of simple stellar population models, and based on those predicted an increase of the $\alpha$-parameter with galaxy colour, in agreement with the observations available at the time. 

In the past two decades, PNe have increasingly been used as kinematic tracers in early-type galaxies as well as to trace the ICL and IGL. Here, I want to especially focus on the transition from ``in-situ'' to accreted (``ex-situ'') stellar populations, which is not only observationally signalled by radially varying light profiles and kinematics but also by changes in the PN population properties. In particular, several studies have found that the $\alpha$-parameter increases significantly from the inner to the outer halos of the group- or cluster-dominating galaxy to the surrounding IGL or ICL \citep[e.g.][]{doherty_2009A&A...502..771D, longobardi_2013A&A...558A..42L, longobardi_2018, hartke_2017A&A...603A.104H, hartke_2020A&A...642A..46H}. Furthermore, motivated by the observed steeper faint-end slope of the PNLF of Virgo ICL-PNe, \citet{longobardi_2013A&A...558A..42L} proposed the modification of eq.~\eqref{eq:PNLF} to allow a variable faint-end slope $c_2$, which has since been applied to observations of PNe in nearby groups and clusters \citep{hartke_2017A&A...603A.104H, hartke_2020A&A...642A..46H} as well as in the nearby M31 \citep{bhattacharya_2021A&A...647A.130B}. While it is difficult to constrain the stellar population properties of the low-surface-brightness haloes, and the surrounding ICL or IGL, pencil-beam surveys suggest that high $\alpha$-parameters and steep PNLF slopes can be linked with old and metal-poor stellar populations \citep[][]{williamsMetallicityDistributionIntracluster2007, harrisLeoEllipticalNGC2007, leeDualStellarHalos2016}, but with the caveat that these measurements are not exactly co-spatial. To interpret the studies of PN populations in groups and clusters of galaxies, it is important to obtain co-spatial measurements of PN and stellar populations, which will be discussed in turn.

\subsection{New insights from integral-field spectroscopic surveys}
\label{ssec:ifus}
Since the pioneering integral-field spectroscopic studies of \citet{roth_2004ApJ...603..531R} and \citet{sarzi_2011MNRAS.415.2832S}, the Multi Unit Spectroscopic Explorer \citep[MUSE,][]{2010SPIE.7735E..08B} has been transformational \citep[see also the review by][]{2023arXiv231114230R} for the study of extragalactic PN populations, especially in late-type galaxies that have been targeted less with classical methods. Furthermore, integral-field spectroscopic observations allow the simultaneous investigation of stellar population properties and discovery and characterisation of PN populations, which are important for putting the results presented in the previous subsection into context.

\citet{kreckel_2017ApJ...834..174K} showed the importance of full spectral information when determining the PNLF distance to NGC~628, as supernova remnants biased previous distance determinations based on narrowband imaging alone \citep{2008ApJ...683..630H} to a shorter distance, despite the MUSE data only covering parts of the disk of the galaxy. 
With a much higher filling factor, the PHANGS-MUSE survey \citep[Physics at High Angular resolution in Nearby GalaxieS;][]{emsellemPHANGSMUSESurveyProbing2022} has been a treasure trove for studying ionized nebulae in nearby star-forming galaxies, with \citet{scheuermann_2022MNRAS.511.6087S} determining PNLF distances to the $19$ galaxies in the sample and can reconcile their derived distances with literature tip of the red-giant branch distances without the need for a metallicity-dependent $M^{\star}$. 
To facilitate the classification of ten thousands of nebulae in the $19$ galaxies into \ion{H}{ii} regions, supernova remnants, and PNe, \citet{2023A&A...672A.148C} developed a Bayesian algorithm, but noted that the PNe sample is incomplete compared to that of \citet{scheuermann_2022MNRAS.511.6087S}, attributing the differences to different source-detection methods, but finding consistent classifications for the sources that are in both catalogues. Building on the PHANGS-MUSE strategy, \citet{congiu_2025A&A...700A.125C} built the largest MUSE mosaic to date, covering the Sculptor galaxy (NGC~253), and resulting in the detection of $\sim500$ PNe. This work demonstrates the importance of accounting for dust in the interstellar medium, which can bias the PNLF distance to larger values. 

Different techniques have been developed to detect PNe from integral-field spectroscopic data, starting with the `classical' visual inspection and blinking of on-off images \citep[e.g.][]{roth_2018A&A...618A...3R}.
\citet{spriggs_2020A&A...637A..62S} developed a technique based on PSF and pixel-by-pixel spectral fitting to automatically detect PNe from Fornax3D survey data targeting the brightest early-type galaxies in the Fornax cluster \citep{2018A&A...616A.121S}. \citet{galan-deanta_2021A&A...652A.109G} applied this technique to three galaxies from Fornax3D to probe metallicity-dependent variations of the $\alpha$-parameter in the centres and (inner) haloes of the galaxies, but did not find any evidence for an increase in the $\alpha$-parameter between metal-rich and metal-poor regions. This differs from the results presented in Section~\ref{ssec:classical} on halo and ICL $\alpha$-parameter variations with stellar populations on significantly larger spatial scales that allowed for probing lower host stellar population metallicities. 

\citet{rothPrecisionCosmologyImproved2021} developed the so-called differential emission line filtering (DELF) method with the goal of facilitating photometric measurements precise enough to use PNLF distances to alleviate the Hubble tension. \citet{jacoby_2024ApJS..271...40J} applied this method to a heterogeneous sample of 20 galaxies with MUSE archival data, demonstrating that the method can yield excellent PNLFs and outline the way forward for PNLF distance measurements, both from an observational standpoint, as well as regarding the need for a better understanding and definition of the analytical form of the PNLF. 

The aforementioned studies were all carried out with the MUSE integral-field spectrograph, that has a prohibitively small field-of-view (FoV, $1^\prime\times1^\prime$ in wide-field mode) to study galaxies in the local Universe in their full extent in a single pointing. The SITELLE instrument \citep[Spectromètre Imageur à Transformée de Fourier pour l'Étude en Long et en Large des raies d'Émission;][]{2012SPIE.8446E..0UG}, an optical imaging Fourier transform spectrometer, covers an $11^\prime\times11^\prime$ FoV, comparable to that of the PN.S instrument. However, the larger FoV is compromised by a shorter wavelength range that can be sampled in one observation. The ongoing Star formation, Ionized Gas, and Nebular Abundances Legacy Survey \citep[SIGNALS;][]{rousseau-nepton_signals_2019} provides observations of $\approx40$ late-type galaxies in the local Universe in three filters, SN1 ($3650-3850$ \AA, covering [\ion{O}{ii}]$\,\lambda 3727$~\AA), SN2 ($4800-5200$ \AA, covering the $\lambda\lambda 4959, 5007$~\AA\ doublet and H$\beta$) and SN3 ($6510-6850$ \AA, covering H$\alpha$ as well as the [\ion{Ne}{ii}]$\lambda\lambda 6548, 6583$~\AA\ and [\ion{S}{ii}]$\lambda\lambda 6716, 6731$~\AA\ lines), allowing for the detection and classification of extragalactic PNe in these galaxies, as piloted by \citet{martin_m31_2018} and \citet{vicens-mouret_2023MNRAS.524.3623V}. 

\subsection{New facilities and opportunities}
\label{sec:newinstr}
The coming years and decades will see the arrival of several telescopes and instruments that may open new and exciting discovery spaces for PN populations outside of the Milky Way. BlueMUSE \citep{2019arXiv190601657R} at the Very Large Telescope, will have a similar design to MUSE that has been transformative for the study of PN populations in nearby galaxies, but covering important bluer emission lines (for example [\ion{O}{ii}]$\,\lambda 3727$~\AA, H$\gamma\,\lambda4340$~\AA, [\ion{O}{iii}]$\,\lambda 4364$~\AA, and \ion{He}{ii}\,$\lambda4686$~\AA\footnote{This line can already be observed with MUSE in the extended spectral configuration, but the majority of extant archival data has been observed in the nominal configuration, starting at 4800~\AA.}). The bluer [\ion{O}{iii}] line is crucial for direct abundance determinations and the \ion{He}{ii} and H$\gamma$ lines provide important constraints for the determination of the excitation class of PNe \citep[e.g.][]{reid_2010PASA...27..187R, bhattacharya_2019A&A...631A..56B}. 

Integral-field units will be complemented by spectroscopic facilities with high multiplexing capabilities and large fields of view, such as the Maunakea Spectroscopic Explorer \citep{2023AN....34430108S} and the Wide-field Spectroscopic Telescope \citep{2024arXiv240305398M} at 10m-class facilities in the northern and southern hemispheres. These will be transformative for the study of individual nebulae, the relation between PNe and their host stellar populations, but also for PNe as tracers of galaxy, halo, and ICL assembly. 

While the bulk of instrumentation for the upcoming \textit{Extremely Large Telescopes} will focus on the infrared wavelength range, several spectrographs will still operate in the optical, covering the important [\ion{O}{iii}]$\lambda\lambda 4959, 5007$~\AA\ doublet from redshift $z=0$, facilitating unprecedentedly deep surveys for PNe in the local universe, as well as pushing the redshift boundaries of extragalactic PN populations beyond the Coma Cluster. The first light for the first generation of these instruments, the High Angular Resolution Monolithic Optical and Near-infrared Integral field spectrograph \citep[HARMONI,][]{2016SPIE.9908E..1XT} at ESO's Extremely Large Telescope (ELT) and the Giant Magellan Telescope Multi-object Astronomical and Cosmological Spectrograph \citep[GMACS,][]{2025PASP..137c5002F} is foreseen for the coming decades. Instruments such as the Multi-AO Imaging Camera for Deep Observations \citep[MICADO,][]{2024SPIE13096E..11S} at the ELT may furthermore complement existing PN surveys with resolved stellar population information for galaxies beyond 10~Mpc. 

\section{Modelling PN populations and insights from stellar evolution models}
\label{sec:models}
The reconciliation of the observed invariance of the absolute magnitude of the PNLF bright cut-off for galaxies of different morphological types with the predictions from stellar evolution models has been a long-standing issue. Furthermore, the use of the PNLF as a distance indicator requires a better theoretical understanding about the PNe that populate the bright end of the PNLF and their origin.
To reconcile observed PN properties with stellar evolution models, one needs to rely on synthetic post-AGB evolution tracks that describe the evolution of low- to intermediate-mass stars following the AGB phase. Historically, two model grids were widely used: \citet{1994ApJS...92..125V} and \citet{1995A&A...299..755B}. While it was possible to reproduce the observed bright cut-off of the PNLF $M^{\star}$ in galaxies with recent star-formation, the presence of similarly bright PNe in elliptical galaxies with old stellar populations could not be reconciled \citep[e.g.][]{marigo_2004A&A...423..995M, ciardullo_2012Ap&SS.341..151C}. 

Furthermore, based on observations of PNe in the Galactic bulge, \citet{gesicki_2014A&A...566A..48G} found that the \citet{1995A&A...299..755B} tracks\footnote{The same holds for the tracks of \citet{1994ApJS...92..125V}, which evolve on even slower timescales.} evolve too slowly to reconcile the predicted with the observed local white dwarf masses.   
The post-AGB evolutionary tracks of \citet{millerbertolami_2016A&A...588A..25M} address this issue, including updated descriptions of micro- and macrophysics, resulting in $3-10\times$ faster post-AGB evolution timescales and overall brighter luminosities. \citet{gesicki_2018} were the first to show that with these faster evolving tracks, even populations with ages between 3 and 7 Gyrs could produce PNe that reach the bright cut-off $M^{\star}$ under the assumption that the brightest PNe are optically thick (maximum-nebula hypothesis). However, combining the new post-AGB tracks \citep{millerbertolami_2016A&A...588A..25M} with the PN population modelling prescription of \citet{1997A&A...321..898M}, \citet{valenzuela_2019ApJ...887...65V} showed that under the maximum-nebula hypothesis too few PNe are produced overall, in conflict with observed PNLFs, and that accounting for optically-thin PNe is important. 

The majority of the above models (and others in the literature) use recipes based on solar measurements and abundances, or, on larger scales, values derived based on Milky Way properties. The PICS (PNe in cosmological simulations) framework \citep{valenzuela_2025A&A...699A.371V} overcomes this limitation by modelling PNe for SSPs with different masses, ages, and metallicities. The PICS models furthermore explore the critical role of dust, using the empirical prescription of \citet{jacoby_2025ApJ...983..129J}, as well as the effect of different prescriptions for the initial-to-final mass relation (IFMR) and for the Helium abundances. While the authors reproduce the general trend from previous models that older SSPs produce less luminous PNe, they also demonstrate that metallicity plays an important role: old SSPs with higher metallicities are able to produce brighter PNe. Furthermore, they find the abundance of the bright PNe to be especially sensitive to the IFMR in old stellar populations, with a flatter IFMR \citep[e.g.][]{2018ApJ...866...21C} leading to larger core masses and thus brighter PNe, alleviating some of the long-standing tension between observations and models, as also discussed in \citet{jacoby_2025ApJ...983..129J}. 

By producing models normalised to the bolometric luminosity, \citet{valenzuela_2025A&A...699A.371V}  also determine the $\alpha$-parameter for a given SSP in a given magnitude range. To facilitate the comparison with the predictions from \citet{buzzoni_2006MNRAS.368..877B}, $\alpha$-parameters are calculated 8 magnitudes from a fixed $M^\star=-4.5$. While the general trends of $\alpha_8$ with age and metallicity are similar between the two models, the use of \citeauthor{millerbertolami_2016A&A...588A..25M}'s \citeyearpar{millerbertolami_2016A&A...588A..25M} metallicity-dependent stellar evolution models by \citet{valenzuela_2025A&A...699A.371V} leads to a stronger dependence of $\alpha_8$ on metallicity. To facilitate the comparison with observations, \citet{valenzuela_2025A&A...699A.371V} also provide predictions for $\alpha_{2.5}$, reducing the need for observers with magnitude-limited data to make assumptions about the shape of the PNLF for extrapolating $\alpha$, and better taking into account the effect of metallicity on the amount of bright PNe. 

\section{Conclusions and outlook}
\label{sec:conclusion}
In this review, I discussed advances of linking extragalactic PN populations with the underlying stellar population properties both from an observational (Sect.~\ref{sec:obs}) and modelling (Sect.~\ref{sec:models}) perspective. Especially on large spatial scales (i.e. comparing galaxy centres and extended haloes or the surrounding ICL in massive environments), there is evidence for a significant change of the $\alpha$-parameter and PNLF shape. However, these measurements are limited by indirect metallicity inferences. To reconcile observations and theory, systematic surveys where PN and stellar population properties are simultaneously measured are needed, covering the entire mass-metallicity relation. As discussed in Section~\ref{ssec:ifus}, integral-field spectrographs open up a new discovery space for PNe and allow for the co-spatial measurement of stellar population properties such as ages, metallicities, and abundances via spectral fitting. 

Ongoing surveys, as well as those planned with new instrumentation at 8-, 10- and 30-m-class telescopes, using integral-field spectroscopy and `classical' PN detection techniques, will provide constraints on the variation of the important diagnostics $\alpha$-parameter and PNLF in stellar populations of in-situ and ex-situ origins. For the interpretation of these observations, but also to further constrain theoretical models of late stellar evolution, it is crucial to properly model extragalactic PN populations in cosmological simulations of galaxy evolution, which has finally become possible, as reviewed in Section~\ref{sec:models}. 

In summary, there are promising studies suggesting that, in the coming years, PNe may be elevated to stellar population tracers in low-surface brightness regions such as galaxy haloes and the ICL -- where stellar population parameters such as age and metallicity cannot be easily measured directly -- in addition to their important role as kinematic tracers. This is fuelled by advances in instrumentation as well as new models of late stellar evolution that will be combined with state-of-the-art numerical simulations of galaxy evolution. 

%TC:ignore
\bibliographystyle{Frontiers-Harvard} % style aa.bst
\bibliography{PNe}

\begin{thebibliography}{72}
\providecommand{\natexlab}[1]{#1}
\expandafter\ifx\csname urlstyle\endcsname\relax
  \providecommand{\doi}[1]{doi:\discretionary{}{}{}#1}\else
  \providecommand{\doi}{doi:\discretionary{}{}{}\begingroup
  \urlstyle{rm}\Url}\fi
\providecommand{\selectlanguage}[1]{\relax}
\providecommand{\bibAnnoteFile}[1]{%
  \IfFileExists{#1}{\begin{quotation}\noindent\textsc{Key:} #1\\
  \textsc{Annotation:}\ \input{#1}\end{quotation}}{}}
\providecommand{\bibAnnote}[2]{%
  \begin{quotation}\noindent\textsc{Key:} #1\\
  \textsc{Annotation:}\ #2\end{quotation}}

\bibitem[{{Aniyan} et~al.(2018){Aniyan}, {Freeman}, {Arnaboldi}, {Gerhard},
  {Coccato}, {Fabricius} et~al.}]{2018MNRAS.476.1909A}
{Aniyan}, S., {Freeman}, K.~C., {Arnaboldi}, M., {Gerhard}, O.~E., {Coccato},
  L., {Fabricius}, M., et~al. (2018).
\newblock {Resolving the disc-halo degeneracy - I: a look at NGC 628}.
\newblock \emph{Monthly Notices of the Royal Astronomical Society} 476,
  1909--1930.
\newblock \doi{10.1093/mnras/sty310}
\bibAnnoteFile{2018MNRAS.476.1909A}

\bibitem[{{Aniyan} et~al.(2021){Aniyan}, {Ponomareva}, {Freeman}, {Arnaboldi},
  {Gerhard}, {Coccato} et~al.}]{2021MNRAS.500.3579A}
{Aniyan}, S., {Ponomareva}, A.~A., {Freeman}, K.~C., {Arnaboldi}, M.,
  {Gerhard}, O.~E., {Coccato}, L., et~al. (2021).
\newblock {Resolving the Disc-Halo Degeneracy - II: NGC 6946}.
\newblock \emph{Monthly Notices of the Royal Astronomical Society} 500,
  3579--3593.
\newblock \doi{10.1093/mnras/staa3106}
\bibAnnoteFile{2021MNRAS.500.3579A}

\bibitem[{Arnaboldi et~al.(1994)Arnaboldi, Freeman, Hui, Capaccioli, and
  Ford}]{arnaboldi_1994Msngr..76...40A}
Arnaboldi, M., Freeman, K.~C., Hui, X., Capaccioli, M., and Ford, H. (1994).
\newblock The kinematics of the planetary nebulae in the outer regions of
  {{NGC}} 1399.
\newblock \emph{The Messenger} 76, 40--44
\bibAnnoteFile{arnaboldi_1994Msngr..76...40A}

\bibitem[{Arnaboldi and
  Gerhard(2022)}]{arnaboldiKinematicsDiffuseIntragroup2022}
Arnaboldi, M. and Gerhard, O. (2022).
\newblock Kinematics of the diffuse intragroup and intracluster light in groups
  and clusters of galaxies in the local universe within 100 {{Mpc}} distance.
\newblock \emph{Frontiers in Astronomy and Space Sciences} 9, 872283.
\newblock \doi{10.3389/fspas.2022.872283}
\bibAnnoteFile{arnaboldiKinematicsDiffuseIntragroup2022}

\bibitem[{{Bacon} et~al.(2010){Bacon}, {Accardo}, {Adjali}, {Anwand}, {Bauer},
  {Biswas} et~al.}]{2010SPIE.7735E..08B}
{Bacon}, R., {Accardo}, M., {Adjali}, L., {Anwand}, H., {Bauer}, S., {Biswas},
  I., et~al. (2010).
\newblock {The MUSE second-generation VLT instrument}.
\newblock In \emph{Ground-based and Airborne Instrumentation for Astronomy
  III}, eds. I.~S. {McLean}, S.~K. {Ramsay}, and H.~{Takami}. vol. 7735 of
  \emph{Society of Photo-Optical Instrumentation Engineers (SPIE) Conference
  Series}, 773508.
\newblock \doi{10.1117/12.856027}
\bibAnnoteFile{2010SPIE.7735E..08B}

\bibitem[{Bhattacharya et~al.(2019)Bhattacharya, Arnaboldi, Caldwell, Gerhard,
  Bla{\~n}a, McConnachie et~al.}]{bhattacharya_2019A&A...631A..56B}
Bhattacharya, S., Arnaboldi, M., Caldwell, N., Gerhard, O., Bla{\~n}a, M.,
  McConnachie, A., et~al. (2019).
\newblock The survey of planetary nebulae in {{Andromeda}} ({{M}} 31). {{II}}.
  {{Age-velocity}} dispersion relation in the disc from planetary nebulae.
\newblock \emph{Astronomy and Astrophysics} 631, A56.
\newblock \doi{10.1051/0004-6361/201935898}
\bibAnnoteFile{bhattacharya_2019A&A...631A..56B}

\bibitem[{Bhattacharya et~al.(2021)Bhattacharya, Arnaboldi, Gerhard,
  McConnachie, Caldwell, Hartke et~al.}]{bhattacharya_2021A&A...647A.130B}
Bhattacharya, S., Arnaboldi, M., Gerhard, O., McConnachie, A., Caldwell, N.,
  Hartke, J., et~al. (2021).
\newblock The survey of planetary nebulae in {{Andromeda}} ({{M}} 31). {{III}}.
  {{Constraints}} from deep planetary nebula luminosity functions on the origin
  of the inner halo substructures in {{M}} 31.
\newblock \emph{Astronomy and Astrophysics} 647, A130.
\newblock \doi{10.1051/0004-6361/202038366}
\bibAnnoteFile{bhattacharya_2021A&A...647A.130B}

\bibitem[{{Bloecker}(1995)}]{1995A&A...299..755B}
{Bloecker}, T. (1995).
\newblock {Stellar evolution of low- and intermediate-mass stars. II. Post-AGB
  evolution.}
\newblock \emph{Astronomy and Astrophysics} 299, 755
\bibAnnoteFile{1995A&A...299..755B}

\bibitem[{Buzzoni(1989)}]{buzzoni_1989ApJS...71..817B}
Buzzoni, A. (1989).
\newblock Evolutionary {{Population Synthesis}} in {{Stellar Systems}}. {{I}}.
  {{A Global Approach}}.
\newblock \emph{The Astrophysical Journal Supplement Series} 71, 817.
\newblock \doi{10.1086/191399}
\bibAnnoteFile{buzzoni_1989ApJS...71..817B}

\bibitem[{Buzzoni et~al.(2006)Buzzoni, Arnaboldi, and
  Corradi}]{buzzoni_2006MNRAS.368..877B}
Buzzoni, A., Arnaboldi, M., and Corradi, R. L.~M. (2006).
\newblock Planetary nebulae as tracers of galaxy stellar populations.
\newblock \emph{Monthly Notices of the Royal Astronomical Society} 368,
  877--894.
\newblock \doi{10.1111/j.1365-2966.2006.10163.x}
\bibAnnoteFile{buzzoni_2006MNRAS.368..877B}

\bibitem[{Ciardullo(2012)}]{ciardullo_2012Ap&SS.341..151C}
Ciardullo, R. (2012).
\newblock The {{Planetary Nebula Luminosity Function}} at the dawn of {{Gaia}}.
\newblock \emph{Astrophysics and Space Science} 341, 151--161.
\newblock \doi{10.1007/s10509-012-1061-2}
\bibAnnoteFile{ciardullo_2012Ap&SS.341..151C}

\bibitem[{Ciardullo(2022)}]{ciardullo_2022}
Ciardullo, R. (2022).
\newblock The {{Planetary Nebula Luminosity Function}} in the {{Era}} of
  {{Precision Cosmology}}.
\newblock \emph{Frontiers in Astronomy and Space Sciences} 9, 896326.
\newblock \doi{10.3389/fspas.2022.896326}
\bibAnnoteFile{ciardullo_2022}

\bibitem[{Ciardullo et~al.(2002)Ciardullo, Feldmeier, Jacoby, de~Naray,
  Laychak, and Durrell}]{ciardullo_2002}
Ciardullo, R., Feldmeier, J.~J., Jacoby, G.~H., de~Naray, R.~K., Laychak,
  M.~B., and Durrell, P.~R. (2002).
\newblock Planetary {{Nebulae}} as {{Standard Candles}}. {{XII}}.
  {{Connecting}} the {{Population I}} and {{Population II Distance Scales}}.
\newblock \emph{The Astrophysical Journal} 577, 31--50.
\newblock \doi{10.1086/342180}
\bibAnnoteFile{ciardullo_2002}

\bibitem[{Ciardullo et~al.(1989)Ciardullo, Jacoby, Ford, and
  Neill}]{ciardullo_1989ApJ...339...53C}
Ciardullo, R., Jacoby, G.~H., Ford, H.~C., and Neill, J.~D. (1989).
\newblock Planetary {{Nebulae}} as {{Standard Candles}}. {{II}}. {{The
  Calibration}} in {{M31}} and {{Its Companions}}.
\newblock \emph{The Astrophysical Journal} 339, 53.
\newblock \doi{10.1086/167275}
\bibAnnoteFile{ciardullo_1989ApJ...339...53C}

\bibitem[{Coccato et~al.(2009)Coccato, Gerhard, Arnaboldi, Das, Douglas,
  Kuijken et~al.}]{coccato_2009MNRAS.394.1249C}
Coccato, L., Gerhard, O., Arnaboldi, M., Das, P., Douglas, N.~G., Kuijken, K.,
  et~al. (2009).
\newblock Kinematic properties of early-type galaxy haloes using planetary
  nebulae*.
\newblock \emph{Monthly Notices of the Royal Astronomical Society} 394,
  1249--1283.
\newblock \doi{10.1111/j.1365-2966.2009.14417.x}
\bibAnnoteFile{coccato_2009MNRAS.394.1249C}

\bibitem[{{Congiu} et~al.(2023){Congiu}, {Blanc}, {Belfiore}, {Santoro},
  {Scheuermann}, {Kreckel} et~al.}]{2023A&A...672A.148C}
{Congiu}, E., {Blanc}, G.~A., {Belfiore}, F., {Santoro}, F., {Scheuermann}, F.,
  {Kreckel}, K., et~al. (2023).
\newblock {PHANGS-MUSE: Detection and Bayesian classification of 40 000 ionised
  nebulae in nearby spiral galaxies}.
\newblock \emph{Astronomy and Astrophysics} 672, A148.
\newblock \doi{10.1051/0004-6361/202245153}
\bibAnnoteFile{2023A&A...672A.148C}

\bibitem[{Congiu et~al.(2025)Congiu, Scheuermann, Kreckel, Leroy, Emsellem,
  Belfiore et~al.}]{congiu_2025A&A...700A.125C}
Congiu, E., Scheuermann, F., Kreckel, K., Leroy, A., Emsellem, E., Belfiore,
  F., et~al. (2025).
\newblock The {{MUSE}} view of the {{Sculptor}} galaxy: {{Survey}} overview and
  the luminosity function of planetary nebulae.
\newblock \emph{Astronomy and Astrophysics} 700, A125.
\newblock \doi{10.1051/0004-6361/202554144}
\bibAnnoteFile{congiu_2025A&A...700A.125C}

\bibitem[{Cortesi et~al.(2013)Cortesi, Arnaboldi, Coccato, Merrifield, Gerhard,
  Bamford et~al.}]{cortesi_2013A&A...549A.115C}
Cortesi, A., Arnaboldi, M., Coccato, L., Merrifield, M.~R., Gerhard, O.,
  Bamford, S., et~al. (2013).
\newblock The {{Planetary Nebula Spectrograph}} survey of {{S0}} galaxy
  kinematics. {{Data}} and overview.
\newblock \emph{Astronomy and Astrophysics} 549, A115.
\newblock \doi{10.1051/0004-6361/201220306}
\bibAnnoteFile{cortesi_2013A&A...549A.115C}

\bibitem[{{Cummings} et~al.(2018){Cummings}, {Kalirai}, {Tremblay},
  {Ramirez-Ruiz}, and {Choi}}]{2018ApJ...866...21C}
{Cummings}, J.~D., {Kalirai}, J.~S., {Tremblay}, P.~E., {Ramirez-Ruiz}, E., and
  {Choi}, J. (2018).
\newblock {The White Dwarf Initial-Final Mass Relation for Progenitor Stars
  from 0.85 to 7.5 M $_{{\ensuremath{\odot}}}$}.
\newblock \emph{The Astrophysical Journal} 866, 21.
\newblock \doi{10.3847/1538-4357/aadfd6}
\bibAnnoteFile{2018ApJ...866...21C}

\bibitem[{Doherty et~al.(2009)Doherty, Arnaboldi, Das, Gerhard, Aguerri,
  Ciardullo et~al.}]{doherty_2009A&A...502..771D}
Doherty, M., Arnaboldi, M., Das, P., Gerhard, O., Aguerri, J. A.~L., Ciardullo,
  R., et~al. (2009).
\newblock The edge of the {{M}} 87 halo and the kinematics of the diffuse light
  in the {{Virgo}} cluster core.
\newblock \emph{Astronomy and Astrophysics} 502, 771--786.
\newblock \doi{10.1051/0004-6361/200811532}
\bibAnnoteFile{doherty_2009A&A...502..771D}

\bibitem[{Dopita et~al.(1992)Dopita, Jacoby, and
  Vassiliadis}]{dopita_1992ApJ...389...27D}
Dopita, M.~A., Jacoby, G.~H., and Vassiliadis, E. (1992).
\newblock A {{Theoretical Calibration}} of the {{Planetary Nebular Cosmic
  Distance Scale}}.
\newblock \emph{The Astrophysical Journal} 389, 27.
\newblock \doi{10.1086/171186}
\bibAnnoteFile{dopita_1992ApJ...389...27D}

\bibitem[{Douglas et~al.(2002)Douglas, Arnaboldi, Freeman, Kuijken, Merrifield,
  Romanowsky et~al.}]{douglas_2002PASP..114.1234D}
Douglas, N.~G., Arnaboldi, M., Freeman, K.~C., Kuijken, K., Merrifield, M.~R.,
  Romanowsky, A.~J., et~al. (2002).
\newblock The {{Planetary Nebula Spectrograph}}: {{The Green Light}} for
  {{Galaxy Kinematics}}.
\newblock \emph{Publications of the Astronomical Society of the Pacific} 114,
  1234--1251.
\newblock \doi{10.1086/342765}
\bibAnnoteFile{douglas_2002PASP..114.1234D}

\bibitem[{Douglas and Taylor(1999)}]{douglas_1999MNRAS.307..190D}
Douglas, N.~G. and Taylor, K. (1999).
\newblock Galaxy kinematics from counter-dispersed imaging.
\newblock \emph{Monthly Notices of the Royal Astronomical Society} 307,
  190--196.
\newblock \doi{10.1046/j.1365-8711.1999.02614.x}
\bibAnnoteFile{douglas_1999MNRAS.307..190D}

\bibitem[{Emsellem et~al.(2022)Emsellem, Schinnerer, Santoro, Belfiore, Pessa,
  McElroy et~al.}]{emsellemPHANGSMUSESurveyProbing2022}
Emsellem, E., Schinnerer, E., Santoro, F., Belfiore, F., Pessa, I., McElroy,
  R., et~al. (2022).
\newblock The {{PHANGS-MUSE}} survey. {{Probing}} the chemo-dynamical evolution
  of disc galaxies.
\newblock \emph{Astronomy and Astrophysics} 659, A191.
\newblock \doi{10.1051/0004-6361/202141727}
\bibAnnoteFile{emsellemPHANGSMUSESurveyProbing2022}

\bibitem[{{Fabricant} et~al.(2025){Fabricant}, {Catropa}, {Fata}, {Brown},
  {Doherty}, {Durusky} et~al.}]{2025PASP..137c5002F}
{Fabricant}, D., {Catropa}, D., {Fata}, R., {Brown}, W., {Doherty}, P.,
  {Durusky}, D., et~al. (2025).
\newblock {GMACS: A Moderate-dispersion Optical Spectrograph for the Giant
  Magellan Telescope}.
\newblock \emph{Publications of the Astronomical Society of the Pacific} 137,
  035002.
\newblock \doi{10.1088/1538-3873/adb0f0}
\bibAnnoteFile{2025PASP..137c5002F}

\bibitem[{Ford and Jenner(1975)}]{ford_1975ApJ...202..365F}
Ford, H.~C. and Jenner, D.~C. (1975).
\newblock Planetary nebulae in local group galaxies. {{II}}.
  {{Identifications}}, positions, number, and production rate of nebulae in
  {{NGC}} 221.
\newblock \emph{The Astrophysical Journal} 202, 365--371.
\newblock \doi{10.1086/153984}
\bibAnnoteFile{ford_1975ApJ...202..365F}

\bibitem[{Ford et~al.(1973)Ford, Jenner, and Epps}]{ford_1973ApJ...183L..73F}
Ford, H.~C., Jenner, D.~C., and Epps, H.~W. (1973).
\newblock Planetary {{Nebulae}} in {{Local-Group Galaxies}}. {{I}}.
  {{Identifications}} in {{NGC}} 185, {{NGC}} 205, and {{NGC}} 221.
\newblock \emph{The Astrophysical Journal} 183, L73.
\newblock \doi{10.1086/181255}
\bibAnnoteFile{ford_1973ApJ...183L..73F}

\bibitem[{{Gal{\'a}n-de Anta} et~al.(2021){Gal{\'a}n-de Anta}, Sarzi, Spriggs,
  Nedelchev, Pinna, {Mart{\'i}n-Navarro}
  et~al.}]{galan-deanta_2021A&A...652A.109G}
{Gal{\'a}n-de Anta}, P.~M., Sarzi, M., Spriggs, T.~W., Nedelchev, B., Pinna,
  F., {Mart{\'i}n-Navarro}, I., et~al. (2021).
\newblock The {{Fornax 3D}} project: {{PNe}} populations and stellar
  metallicity in edge-on galaxies.
\newblock \emph{Astronomy and Astrophysics} 652, A109.
\newblock \doi{10.1051/0004-6361/202140834}
\bibAnnoteFile{galan-deanta_2021A&A...652A.109G}

\bibitem[{Gerhard et~al.(2005)Gerhard, Arnaboldi, Freeman, Kashikawa, Okamura,
  and Yasuda}]{gerhard_2005ApJ...621L..93G}
Gerhard, O., Arnaboldi, M., Freeman, K.~C., Kashikawa, N., Okamura, S., and
  Yasuda, N. (2005).
\newblock Detection of {{Intracluster Planetary Nebulae}} in the {{Coma
  Cluster}}.
\newblock \emph{The Astrophysical Journal} 621, L93--L96.
\newblock \doi{10.1086/429221}
\bibAnnoteFile{gerhard_2005ApJ...621L..93G}

\bibitem[{Gesicki et~al.(2018)Gesicki, Zijlstra, and Bertolami}]{gesicki_2018}
Gesicki, K., Zijlstra, A.~A., and Bertolami, M. M.~M. (2018).
\newblock The mysterious age invariance of the planetary nebula luminosity
  function bright cut-off.
\newblock \emph{Nature Astronomy} 92, 1--5.
\newblock \doi{10.1038/s41550-018-0453-9}
\bibAnnoteFile{gesicki_2018}

\bibitem[{Gesicki et~al.(2014)Gesicki, Zijlstra, Hajduk, and
  Szyszka}]{gesicki_2014A&A...566A..48G}
Gesicki, K., Zijlstra, A.~A., Hajduk, M., and Szyszka, C. (2014).
\newblock Accelerated post-{{AGB}} evolution, initial-final mass relations, and
  the star-formation history of the {{Galactic}} bulge.
\newblock \emph{Astronomy and Astrophysics} 566, A48.
\newblock \doi{10.1051/0004-6361/201118391}
\bibAnnoteFile{gesicki_2014A&A...566A..48G}

\bibitem[{{Grandmont} et~al.(2012){Grandmont}, {Drissen}, {Mandar}, {Thibault},
  and {Baril}}]{2012SPIE.8446E..0UG}
{Grandmont}, F., {Drissen}, L., {Mandar}, J., {Thibault}, S., and {Baril}, M.
  (2012).
\newblock {Final design of SITELLE: a wide-field imaging Fourier transform
  spectrometer for the Canada-France-Hawaii Telescope}.
\newblock In \emph{Ground-based and Airborne Instrumentation for Astronomy IV},
  eds. I.~S. {McLean}, S.~K. {Ramsay}, and H.~{Takami}. vol. 8446 of
  \emph{Society of Photo-Optical Instrumentation Engineers (SPIE) Conference
  Series}, 84460U.
\newblock \doi{10.1117/12.926782}
\bibAnnoteFile{2012SPIE.8446E..0UG}

\bibitem[{Harris et~al.(2007)Harris, Harris, Layden, and
  Wehner}]{harrisLeoEllipticalNGC2007}
Harris, W.~E., Harris, G. L.~H., Layden, A.~C., and Wehner, E. M.~H. (2007).
\newblock The {{Leo Elliptical NGC}} 3379: {{A Metal-Poor Halo Emerges}}.
\newblock \emph{The Astrophysical Journal} 666, 903--918.
\newblock \doi{10.1086/520799}
\bibAnnoteFile{harrisLeoEllipticalNGC2007}

\bibitem[{Hartke et~al.(2020)Hartke, Arnaboldi, Gerhard, Coccato, Pulsoni,
  Freeman et~al.}]{hartke_2020A&A...642A..46H}
Hartke, J., Arnaboldi, M., Gerhard, O., Coccato, L., Pulsoni, C., Freeman,
  K.~C., et~al. (2020).
\newblock The halo of {{M}} 105 and its group environment as traced by
  planetary nebula populations. {{I}}. {{Wide-field}} photometric survey of
  planetary nebulae in the {{Leo I}} group.
\newblock \emph{Astronomy and Astrophysics} 642, A46.
\newblock \doi{10.1051/0004-6361/202038009}
\bibAnnoteFile{hartke_2020A&A...642A..46H}

\bibitem[{Hartke et~al.(2017)Hartke, Arnaboldi, Longobardi, Gerhard, Freeman,
  Okamura et~al.}]{hartke_2017A&A...603A.104H}
Hartke, J., Arnaboldi, M., Longobardi, A., Gerhard, O., Freeman, K.~C.,
  Okamura, S., et~al. (2017).
\newblock The halo of {{M}} 49 and its environment as traced by planetary
  nebulae populations.
\newblock \emph{Astronomy and Astrophysics} 603, A104.
\newblock \doi{10.1051/0004-6361/201730463}
\bibAnnoteFile{hartke_2017A&A...603A.104H}

\bibitem[{Henize and Westerlund(1963)}]{henize_1963ApJ...137..747H}
Henize, K.~G. and Westerlund, B.~E. (1963).
\newblock Dimensions of {{Diffuse}} and {{Planetary Nebulae}} in the {{Small
  Magellanic Cloud}}.
\newblock \emph{The Astrophysical Journal} 137, 747.
\newblock \doi{10.1086/147552}
\bibAnnoteFile{henize_1963ApJ...137..747H}

\bibitem[{{Herrmann} et~al.(2008){Herrmann}, {Ciardullo}, {Feldmeier}, and
  {Vinciguerra}}]{2008ApJ...683..630H}
{Herrmann}, K.~A., {Ciardullo}, R., {Feldmeier}, J.~J., and {Vinciguerra}, M.
  (2008).
\newblock {Planetary Nebulae in Face-On Spiral Galaxies. I. Planetary Nebula
  Photometry and Distances}.
\newblock \emph{The Astrophysical Journal} 683, 630--643.
\newblock \doi{10.1086/589920}
\bibAnnoteFile{2008ApJ...683..630H}

\bibitem[{Hui et~al.(1995)Hui, Ford, Freeman, and
  Dopita}]{hui_1995ApJ...449..592H}
Hui, X., Ford, H.~C., Freeman, K.~C., and Dopita, M.~A. (1995).
\newblock The {{Planetary Nebula System}} and {{Dynamics}} of {{NGC}} 5128.
  {{III}}. {{Kinematics}} and {{Halo Mass Distributions}}.
\newblock \emph{The Astrophysical Journal} 449, 592.
\newblock \doi{10.1086/176082}
\bibAnnoteFile{hui_1995ApJ...449..592H}

\bibitem[{Jacoby(1980)}]{jacoby_1980ApJS...42....1J}
Jacoby, G.~H. (1980).
\newblock The luminosity function for planetary nebulae and the number of
  planetary nebulae in local group galaxies.
\newblock \emph{The Astrophysical Journal Supplement Series} 42, 1--18.
\newblock \doi{10.1086/190642}
\bibAnnoteFile{jacoby_1980ApJS...42....1J}

\bibitem[{Jacoby(1989)}]{jacoby_1989ApJ...339...39J}
Jacoby, G.~H. (1989).
\newblock Planetary {{Nebulae}} as {{Standard Candles}}. {{I}}. {{Evolutionary
  Models}}.
\newblock \emph{The Astrophysical Journal} 339, 39.
\newblock \doi{10.1086/167274}
\bibAnnoteFile{jacoby_1989ApJ...339...39J}

\bibitem[{Jacoby and Ciardullo(2025)}]{jacoby_2025ApJ...983..129J}
Jacoby, G.~H. and Ciardullo, R. (2025).
\newblock The {{Critical Role}} of {{Dust}} on the [{{O III}}] {{Planetary
  Nebula Luminosity Function}}'s {{Bright-end Cutoff}}.
\newblock \emph{The Astrophysical Journal} 983, 129.
\newblock \doi{10.3847/1538-4357/adc0fb}
\bibAnnoteFile{jacoby_2025ApJ...983..129J}

\bibitem[{Jacoby et~al.(2024)Jacoby, Ciardullo, Roth, Arnaboldi, and
  Weilbacher}]{jacoby_2024ApJS..271...40J}
Jacoby, G.~H., Ciardullo, R., Roth, M.~M., Arnaboldi, M., and Weilbacher, P.~M.
  (2024).
\newblock Toward {{Precision Cosmology}} with {{Improved Planetary Nebula
  Luminosity Function Distances Using VLT-MUSE}}. {{II}}. {{A Test Sample}}
  from {{Archival Data}}.
\newblock \emph{The Astrophysical Journal Supplement Series} 271, 40.
\newblock \doi{10.3847/1538-4365/ad2166}
\bibAnnoteFile{jacoby_2024ApJS..271...40J}

\bibitem[{Kreckel et~al.(2017)Kreckel, Groves, Bigiel, Blanc, Kruijssen, Hughes
  et~al.}]{kreckel_2017ApJ...834..174K}
Kreckel, K., Groves, B., Bigiel, F., Blanc, G.~A., Kruijssen, J. M.~D., Hughes,
  A., et~al. (2017).
\newblock A {{Revised Planetary Nebula Luminosity Function Distance}} to
  {{NGC}} 628 {{Using MUSE}}.
\newblock \emph{The Astrophysical Journal} 834, 174.
\newblock \doi{10.3847/1538-4357/834/2/174}
\bibAnnoteFile{kreckel_2017ApJ...834..174K}

\bibitem[{Lee and Jang(2016)}]{leeDualStellarHalos2016}
Lee, M.~G. and Jang, I.~S. (2016).
\newblock Dual {{Stellar Halos}} in the {{Standard Elliptical Galaxy M105}} and
  {{Formation}} of {{Massive Early-type Galaxies}}.
\newblock \emph{The Astrophysical Journal} 822, 70.
\newblock \doi{10.3847/0004-637X/822/2/70}
\bibAnnoteFile{leeDualStellarHalos2016}

\bibitem[{Longobardi et~al.(2013)Longobardi, Arnaboldi, Gerhard, Coccato,
  Okamura, and Freeman}]{longobardi_2013A&A...558A..42L}
Longobardi, A., Arnaboldi, M., Gerhard, O., Coccato, L., Okamura, S., and
  Freeman, K.~C. (2013).
\newblock The planetary nebula population in the halo of {{M}} 87.
\newblock \emph{Astronomy and Astrophysics} 558, A42.
\newblock \doi{10.1051/0004-6361/201321652}
\bibAnnoteFile{longobardi_2013A&A...558A..42L}

\bibitem[{Longobardi et~al.(2018)Longobardi, Arnaboldi, Gerhard, Pulsoni, and
  {S{\"o}ldner-Rembold}}]{longobardi_2018}
Longobardi, A., Arnaboldi, M., Gerhard, O., Pulsoni, C., and
  {S{\"o}ldner-Rembold}, I. (2018).
\newblock Kinematics of the outer halo of {{M}} 87 as mapped by planetary
  nebulae{$\star$}.
\newblock \emph{Astronomy \& Astrophysics} 620, A111.
\newblock \doi{10.1051/0004-6361/201832729}
\bibAnnoteFile{longobardi_2018}

\bibitem[{{Mainieri} et~al.(2024){Mainieri}, {Anderson}, {Brinchmann},
  {Cimatti}, {Ellis}, {Hill} et~al.}]{2024arXiv240305398M}
{Mainieri}, V., {Anderson}, R.~I., {Brinchmann}, J., {Cimatti}, A., {Ellis},
  R.~S., {Hill}, V., et~al. (2024).
\newblock {The Wide-field Spectroscopic Telescope (WST) Science White Paper}.
\newblock \emph{arXiv e-prints} ,
  arXiv:2403.05398\doi{10.48550/arXiv.2403.05398}
\bibAnnoteFile{2024arXiv240305398M}

\bibitem[{Marigo et~al.(2004)Marigo, Girardi, Weiss, Groenewegen, and
  Chiosi}]{marigo_2004A&A...423..995M}
Marigo, P., Girardi, L., Weiss, A., Groenewegen, M. A.~T., and Chiosi, C.
  (2004).
\newblock Evolution of planetary nebulae. {{II}}. {{Population}} effects on the
  bright cut-off of the {{PNLF}}.
\newblock \emph{Astronomy and Astrophysics} 423, 995--1015.
\newblock \doi{10.1051/0004-6361:20040234}
\bibAnnoteFile{marigo_2004A&A...423..995M}

\bibitem[{Martin et~al.(2018)Martin, Drissen, and Melchior}]{martin_m31_2018}
Martin, T.~B., Drissen, L., and Melchior, A.-L. (2018).
\newblock A {SITELLE} view of {M31}'s central region - {I}. {Calibrations} and
  radial velocity catalogue of nearly 800 emission-line point-like sources.
\newblock \emph{Monthly Notices of the Royal Astronomical Society} 473,
  4130--4149.
\newblock \doi{10.1093/mnras/stx2513}.
\newblock ADS Bibcode: 2018MNRAS.473.4130M
\bibAnnoteFile{martin_m31_2018}

\bibitem[{{Mendez} and {Soffner}(1997)}]{1997A&A...321..898M}
{Mendez}, R.~H. and {Soffner}, T. (1997).
\newblock {Improved simulations of the planetary nebula luminosity function.}
\newblock \emph{Astronomy and Astrophysics} 321, 898--906.
\newblock \doi{10.48550/arXiv.astro-ph/9611128}
\bibAnnoteFile{1997A&A...321..898M}

\bibitem[{Miller~Bertolami(2016)}]{millerbertolami_2016A&A...588A..25M}
Miller~Bertolami, M.~M. (2016).
\newblock New models for the evolution of post-asymptotic giant branch stars
  and central stars of planetary nebulae.
\newblock \emph{Astronomy and Astrophysics} 588, A25.
\newblock \doi{10.1051/0004-6361/201526577}
\bibAnnoteFile{millerbertolami_2016A&A...588A..25M}

\bibitem[{Pulsoni et~al.(2018)Pulsoni, Gerhard, Arnaboldi, Coccato, Longobardi,
  Napolitano et~al.}]{pulsoni_2018A&A...618A..94P}
Pulsoni, C., Gerhard, O., Arnaboldi, M., Coccato, L., Longobardi, A.,
  Napolitano, N.~R., et~al. (2018).
\newblock The extended {{Planetary Nebula Spectrograph}} ({{ePN}}.{{S}})
  early-type galaxy survey: {{The}} kinematic diversity of stellar halos and
  the relation between halo transition scale and stellar mass.
\newblock \emph{Astronomy and Astrophysics} 618, A94.
\newblock \doi{10.1051/0004-6361/201732473}
\bibAnnoteFile{pulsoni_2018A&A...618A..94P}

\bibitem[{Reid and Parker(2010)}]{reid_2010PASA...27..187R}
Reid, W.~A. and Parker, Q.~A. (2010).
\newblock An {{Evaluation}} of the {{Excitation-Class Parameter}} for the
  {{Central Stars}} of {{Planetary Nebulae}}.
\newblock \emph{Publications of the Astronomical Society of Australia} 27,
  187--198.
\newblock \doi{10.1071/AS09055}
\bibAnnoteFile{reid_2010PASA...27..187R}

\bibitem[{Renzini and Buzzoni(1986)}]{renziniGlobalPropertiesStellar1986}
Renzini, A. and Buzzoni, A. (1986).
\newblock Global properties of stellar populations and the spectral evolution
  of galaxies (Spectral Evolution of Galaxies), vol. 122 of \emph{Spectral
  {{Evolution}} of {{Galaxies}}}, 195--231.
\newblock \doi{10.1007/978-94-009-4598-2_19}
\bibAnnoteFile{renziniGlobalPropertiesStellar1986}

\bibitem[{{Richard} et~al.(2019){Richard}, {Bacon}, {Blaizot}, {Boissier},
  {Boselli}, {NicolasBouch{\'e}} et~al.}]{2019arXiv190601657R}
{Richard}, J., {Bacon}, R., {Blaizot}, J., {Boissier}, S., {Boselli}, A.,
  {NicolasBouch{\'e}}, et~al. (2019).
\newblock {BlueMUSE: Project Overview and Science Cases}.
\newblock \emph{arXiv e-prints} ,
  arXiv:1906.01657\doi{10.48550/arXiv.1906.01657}
\bibAnnoteFile{2019arXiv190601657R}

\bibitem[{Roth et~al.(2004)Roth, Becker, Kelz, and
  Schmoll}]{roth_2004ApJ...603..531R}
Roth, M.~M., Becker, T., Kelz, A., and Schmoll, J. (2004).
\newblock Spectrophotometry of {{Planetary Nebulae}} in the {{Bulge}} of
  {{M31}}.
\newblock \emph{The Astrophysical Journal} 603, 531--547.
\newblock \doi{10.1086/381526}
\bibAnnoteFile{roth_2004ApJ...603..531R}

\bibitem[{{Roth} et~al.(2023){Roth}, {Jacoby}, {Ciardullo}, {Soemitro},
  {Weilbacher}, and {Arnaboldi}}]{2023arXiv231114230R}
{Roth}, M.~M., {Jacoby}, G., {Ciardullo}, R., {Soemitro}, A., {Weilbacher},
  P.~M., and {Arnaboldi}, M. (2023).
\newblock {Integral Field Spectroscopy: a disruptive innovation for
  observations of Planetary Nebulae and the PNLF}.
\newblock \emph{arXiv e-prints} ,
  arXiv:2311.14230\doi{10.48550/arXiv.2311.14230}
\bibAnnoteFile{2023arXiv231114230R}

\bibitem[{Roth et~al.(2021)Roth, Jacoby, Ciardullo, Davis, Chase, and
  Weilbacher}]{rothPrecisionCosmologyImproved2021}
Roth, M.~M., Jacoby, G.~H., Ciardullo, R., Davis, B.~D., Chase, O., and
  Weilbacher, P.~M. (2021).
\newblock Toward {{Precision Cosmology}} with {{Improved PNLF Distances Using
  VLT-MUSEI}}. {{Methodology}} and {{Tests}}.
\newblock \emph{The Astrophysical Journal} 916, 21.
\newblock \doi{10.3847/1538-4357/ac02ca}
\bibAnnoteFile{rothPrecisionCosmologyImproved2021}

\bibitem[{Roth et~al.(2018)Roth, Sandin, Kamann, Husser, Weilbacher,
  {Monreal-Ibero} et~al.}]{roth_2018A&A...618A...3R}
Roth, M.~M., Sandin, C., Kamann, S., Husser, T.-O., Weilbacher, P.~M.,
  {Monreal-Ibero}, A., et~al. (2018).
\newblock {{MUSE}} crowded field {{3D}} spectroscopy in {{NGC}} 300. {{I}}.
  {{First}} results from central fields.
\newblock \emph{Astronomy and Astrophysics} 618, A3.
\newblock \doi{10.1051/0004-6361/201833007}
\bibAnnoteFile{roth_2018A&A...618A...3R}

\bibitem[{Rousseau-Nepton et~al.(2019)Rousseau-Nepton, Martin, Robert, Drissen,
  Amram, Prunet et~al.}]{rousseau-nepton_signals_2019}
Rousseau-Nepton, L., Martin, R.~P., Robert, C., Drissen, L., Amram, P., Prunet,
  S., et~al. (2019).
\newblock {SIGNALS}: {I}. {Survey} description.
\newblock \emph{Monthly Notices of the Royal Astronomical Society} 489,
  5530--5546.
\newblock \doi{10.1093/mnras/stz2455}.
\newblock ADS Bibcode: 2019MNRAS.489.5530R
\bibAnnoteFile{rousseau-nepton_signals_2019}

\bibitem[{{Sarzi} et~al.(2018){Sarzi}, {Iodice}, {Coccato}, {Corsini}, {de
  Zeeuw}, {Falc{\'o}n-Barroso} et~al.}]{2018A&A...616A.121S}
{Sarzi}, M., {Iodice}, E., {Coccato}, L., {Corsini}, E.~M., {de Zeeuw}, P.~T.,
  {Falc{\'o}n-Barroso}, J., et~al. (2018).
\newblock {Fornax3D project: Overall goals, galaxy sample, MUSE data analysis,
  and initial results}.
\newblock \emph{Astronomy and Astrophysics} 616, A121.
\newblock \doi{10.1051/0004-6361/201833137}
\bibAnnoteFile{2018A&A...616A.121S}

\bibitem[{Sarzi et~al.(2011)Sarzi, Mamon, Cappellari, Emsellem, Bacon, Davies
  et~al.}]{sarzi_2011MNRAS.415.2832S}
Sarzi, M., Mamon, G.~A., Cappellari, M., Emsellem, E., Bacon, R., Davies,
  R.~L., et~al. (2011).
\newblock The planetary nebulae population in the central regions of {{M32}}:
  The {{SAURON}} view.
\newblock \emph{Monthly Notices of the Royal Astronomical Society} 415,
  2832--2843.
\newblock \doi{10.1111/j.1365-2966.2011.18900.x}
\bibAnnoteFile{sarzi_2011MNRAS.415.2832S}

\bibitem[{Scheuermann et~al.(2022)Scheuermann, Kreckel, Anand, Blanc, Congiu,
  Santoro et~al.}]{scheuermann_2022MNRAS.511.6087S}
Scheuermann, F., Kreckel, K., Anand, G.~S., Blanc, G.~A., Congiu, E., Santoro,
  F., et~al. (2022).
\newblock Planetary nebula luminosity function distances for 19 galaxies
  observed by {{PHANGS-MUSE}}.
\newblock \emph{Monthly Notices of the Royal Astronomical Society} 511,
  6087--6109.
\newblock \doi{10.1093/mnras/stac110}
\bibAnnoteFile{scheuermann_2022MNRAS.511.6087S}

\bibitem[{{Sheinis} et~al.(2023){Sheinis}, {Barden}, and
  {Sobeck}}]{2023AN....34430108S}
{Sheinis}, A., {Barden}, S.~C., and {Sobeck}, J. (2023).
\newblock {The Maunakea Spectroscopic Explorer: Thousands of fibers, infinite
  possibilities}.
\newblock \emph{Astronomische Nachrichten} 344, e20230108.
\newblock \doi{10.1002/asna.20230108}
\bibAnnoteFile{2023AN....34430108S}

\bibitem[{Spriggs et~al.(2020)Spriggs, Sarzi, Napiwotzki, {Gal{\'a}n-de Anta},
  Viaene, Nedelchev et~al.}]{spriggs_2020A&A...637A..62S}
Spriggs, T.~W., Sarzi, M., Napiwotzki, R., {Gal{\'a}n-de Anta}, P.~M., Viaene,
  S., Nedelchev, B., et~al. (2020).
\newblock Fornax {{3D}} project: {{Automated}} detection of planetary nebulae
  in the centres of early-type galaxies and first results.
\newblock \emph{Astronomy and Astrophysics} 637, A62.
\newblock \doi{10.1051/0004-6361/201936862}
\bibAnnoteFile{spriggs_2020A&A...637A..62S}

\bibitem[{{Sturm} et~al.(2024){Sturm}, {Davies}, {Alves}, {Cl{\'e}net},
  {Kotilainen}, {Monna} et~al.}]{2024SPIE13096E..11S}
{Sturm}, E., {Davies}, R., {Alves}, J., {Cl{\'e}net}, Y., {Kotilainen}, J.,
  {Monna}, A., et~al. (2024).
\newblock {The MICADO first light imager for the ELT: overview and current
  status}.
\newblock In \emph{Ground-based and Airborne Instrumentation for Astronomy X},
  eds. J.~J. {Bryant}, K.~{Motohara}, and J.~R.~D. {Vernet}. vol. 13096 of
  \emph{Society of Photo-Optical Instrumentation Engineers (SPIE) Conference
  Series}, 1309611.
\newblock \doi{10.1117/12.3017752}
\bibAnnoteFile{2024SPIE13096E..11S}

\bibitem[{{Thatte} et~al.(2016){Thatte}, {Clarke}, {Bryson}, {Shnetler},
  {Tecza}, {Fusco} et~al.}]{2016SPIE.9908E..1XT}
{Thatte}, N.~A., {Clarke}, F., {Bryson}, I., {Shnetler}, H., {Tecza}, M.,
  {Fusco}, T., et~al. (2016).
\newblock {The E-ELT first light spectrograph HARMONI: capabilities and modes}.
\newblock In \emph{Ground-based and Airborne Instrumentation for Astronomy VI},
  eds. C.~J. {Evans}, L.~{Simard}, and H.~{Takami}. vol. 9908 of \emph{Society
  of Photo-Optical Instrumentation Engineers (SPIE) Conference Series}, 99081X.
\newblock \doi{10.1117/12.2230629}
\bibAnnoteFile{2016SPIE.9908E..1XT}

\bibitem[{Valenzuela et~al.(2019)Valenzuela, M{\'e}ndez, and
  Miller~Bertolami}]{valenzuela_2019ApJ...887...65V}
Valenzuela, L.~M., M{\'e}ndez, R.~H., and Miller~Bertolami, M.~M. (2019).
\newblock Revised {{Simulations}} of the {{Planetary Nebulae Luminosity
  Function}}.
\newblock \emph{The Astrophysical Journal} 887, 65.
\newblock \doi{10.3847/1538-4357/ab4e96}
\bibAnnoteFile{valenzuela_2019ApJ...887...65V}

\bibitem[{Valenzuela et~al.(2025)Valenzuela, Miller~Bertolami, Remus, and
  M{\'e}ndez}]{valenzuela_2025A&A...699A.371V}
Valenzuela, L.~M., Miller~Bertolami, M.~M., Remus, R.-S., and M{\'e}ndez, R.~H.
  (2025).
\newblock The {{PICS Project}}: {{I}}. {{The}} impact of metallicity and helium
  abundance on the bright end of the planetary nebula luminosity function.
\newblock \emph{Astronomy and Astrophysics} 699, A371.
\newblock \doi{10.1051/0004-6361/202553974}
\bibAnnoteFile{valenzuela_2025A&A...699A.371V}

\bibitem[{{Vassiliadis} and {Wood}(1994)}]{1994ApJS...92..125V}
{Vassiliadis}, E. and {Wood}, P.~R. (1994).
\newblock {Post--Asymptotic Giant Branch Evolution of Low- to Intermediate-Mass
  Stars}.
\newblock \emph{The Astrophysical Journal Supplement Series} 92, 125.
\newblock \doi{10.1086/191962}
\bibAnnoteFile{1994ApJS...92..125V}

\bibitem[{{Vicens-Mouret} et~al.(2023){Vicens-Mouret}, Drissen, Robert,
  {Rousseau-Nepton}, Martin, and Amram}]{vicens-mouret_2023MNRAS.524.3623V}
{Vicens-Mouret}, S., Drissen, L., Robert, C., {Rousseau-Nepton}, L., Martin,
  R.~P., and Amram, P. (2023).
\newblock Planetary nebulae and supernova remnants in {{NGC}} 4214 from the
  {{SIGNALS}} survey.
\newblock \emph{Monthly Notices of the Royal Astronomical Society} 524,
  3623--3635.
\newblock \doi{10.1093/mnras/stad2154}
\bibAnnoteFile{vicens-mouret_2023MNRAS.524.3623V}

\bibitem[{Williams et~al.(2007)Williams, Ciardullo, Durrell, Vinciguerra,
  Feldmeier, Jacoby et~al.}]{williamsMetallicityDistributionIntracluster2007}
Williams, B.~F., Ciardullo, R., Durrell, P.~R., Vinciguerra, M., Feldmeier,
  J.~J., Jacoby, G.~H., et~al. (2007).
\newblock The {{Metallicity Distribution}} of {{Intracluster Stars}} in
  {{Virgo}}.
\newblock \emph{The Astrophysical Journal} 656, 756--769.
\newblock \doi{10.1086/510149}
\bibAnnoteFile{williamsMetallicityDistributionIntracluster2007}

\end{thebibliography}

\section*{Conflict of Interest Statement}
%All financial, commercial or other relationships that might be perceived by the academic community as representing a potential conflict of interest must be disclosed. If no such relationship exists, authors will be asked to confirm the following statement: 

The author declares that the research was conducted in the absence of any commercial or financial relationships that could be construed as a potential conflict of interest.

\section*{Author Contributions}
JH: conceptualisation, writing -- original draft.

\section*{Funding}
JH acknowledges the support from the Turku Collegium for Science, Medicine and Technology (TCSMT) in the form of a starting grant and the financial support from the Visitor and Mobility program of the Finnish Centre for Astronomy with ESO (FINCA).

\section*{Acknowledgments}
JH thanks her colleagues, in particular 
M. Arnaboldi,
S. Bhattacharya, 
M. Bureau,
L. Coccato,
E. Congiu,
A. Ennis,
O. Gerhard,
C. Pulsoni,
M. Sarzi,
S. Penger, 
C. Spiniello,
L. Valenzuela, and
N. Yang, 
for their contributions and discussions leading to this mini-review and also thanks the anonymous referee for their constructive comments that improved the manuscript. 

%TC:endignore
\end{document}